\documentclass[12pt]{iopart}
\usepackage{color}
\usepackage{bm}
\usepackage{citesort}
\usepackage{graphicx}
\usepackage{amssymb}
\usepackage[dvipsnames]{xcolor}

\def\be{\begin{equation}}
\def\ee{\end{equation}}
\def\bea{\begin{eqnarray}}
\def\eea{\end{eqnarray}}

\def\ptl{\partial}

\usepackage{citesort}

\begin{document}
\nocite{*}

\title{Wave optics of quantum gravity for massive particles}

\noindent \qquad July 20, 2020

\author{S.L. Cherkas\dag   \ and V.L. Kalashnikov\ddag}

\address{\dag\
Institute for Nuclear Problems, Bobruiskaya 11, Minsk 220006,
Belarus}

\address{\ddag\ Facolt\'a
di Ingegneria dell'Informazione, Informatica e Statistica,
Sapienza Universit\'a di Roma, Via Eudossiana 18 00189 - Roma, RM,
Italia}

\begin{abstract} Effects of the quantum gravity under Minkowski space-time background
are considered. It is shown that despite the absence of the complete
theory of quantum gravity, some concrete predictions could be made
for the influence of the quantum gravitational fluctuations on the
propagation of the massive particles. We demonstrate that although
the gravitational potential fluctuations do not produce
particle scattering, they cause decoherence of the matter waves
due to off-shell effects. For point-like massive particles of the
Planck mass order, the effect is considerable. However, this type of decoherence is beyond the measurable possibility for the real particles of the finite size.
\end{abstract}

\section{Introduction}

It is widely stated that the complete theory of quantum gravity (QG)
is not built yet. Indeed, it is true.  At the same time, it is
usually implied that the quantum gravitational fluctuations of
space-time should be small. However, within the theory of general
relativity (GR), one could hardly state that the quantum
gravitational fluctuations are small because the
coordinates' transformation to the reference frame where an observer has a highly oscillating position would result in
substantial quantum gravitational fluctuations. Moreover, a number
of real particles will be created from a vacuum in such a
reference frame \cite{Birrell82}.

The situation changes cardinally when some preferred system of
reference exists. For instance, the cosmic microwave background
(CMB) defines the reference frame where CMB dipole anisotropy is
absent \cite{Dodelson2003}. That suggests considering all the
phenomena in this particular frame. However, the CMB alone is not
sufficient for determining the reference frame uniquely.

Another landmark is the vacuum energy problem insisting and
specifying a class of permitted metrics
\cite{Vesti,nonlin20,Haridasu2020}. As shown,
a conformally-unimodular gauge \cite{Vesti} allows extending the GR
to some theory admitting a Hamiltonian constraint satisfied up to
some constant \cite{Vesti}. That explains why the main part of
vacuum energy $\rho_{vac}\sim M_p^4$ does not contribute to
gravity \cite{Haridasu2020}, i.e., does not lead to the very fast
universe expansion.

Observation of the QG effects in table-top, the accelerator experiments or
astrophysics  is a dream of the several physicist generations
\cite{Power2000,Amel,Wang2012,Albrecht2014,Bassi2017,Carney2019,Diosi2019,howl2020testing}.
Here we will consider the simplest vacuum model as a medium with
the stochastic gravitational potential \cite{Hu2008} and consider
propagating the massive particles through it.

\section{From GR to the gauge violating theory of gravity}

In GR, any spatially uniform energy density (including that of zero-point
fluctuations of the quantum fields) causes the expansion of the
universe. Using the Planck level of the ultra-violet (UV) cutoff of momentum results in the Planckian vacuum energy density $\rho_{vac}\sim M_p^4$ \cite{w},
which must lead to the universe expansion with the Planckian rate
\cite{d1}. In this sense, because such a fast expansion is not
experimentally visible, the vacuum energy problem is an
observational fact \cite{Haridasu2020}. One of the obvious
solutions is to build a theory of gravity, allowing an arbitrarily
reference level of energy density. One such theory has long been
known. That is the unimodular gravity \cite{a17,a21,a22,a23,unim},
which admits an arbitrary cosmological constant. However, under
using the UV comoving momentums cutoff, the vacuum energy density
scales with time as radiation \cite{Visser18,Haridasu2020}, but
not as the cosmological constant.

Another theory \cite{Vesti} could also lead to Friedmann's
equation defined up to some arbitrary constant, but this constant
corresponds to the invisible radiation and can compensate the
vacuum energy. This five-vector theory of gravity (FVT)
\cite{Vesti} assumes the gauge invariance violation of GR  by
constraining the class of all possible metrics in varying the
standard Einstein-Hilbert action. One has to vary not over all
possible space-time metrics $g_{\mu\nu}$, but over some class of
conformally-unimodular metrics
\be
\fl ds^2\equiv g_{\mu\nu} dx^\mu dx^\nu = a^2\left(1-\ptl_m
P^m\right)^2d\eta^2-\gamma_{ij} (dx^i+ N^i d\eta) (dx^j+
N^jd\eta),
\label{interv1}
\ee
where $x^\mu=\{\eta,\bm x\}$, $\eta$ is a conformal time,
$\gamma_{ij}$ is a spatial metric, $a =\gamma^{1/6}$ is a locally
defined scale factor, and $\gamma=\det\gamma_{ij}$. The spatial
part of the interval (\ref{interv1}) reads as
\be
dl^2\equiv\gamma_{ij}dx^idx^j=a^2(\eta,\bm x)\tilde
\gamma_{ij}dx^idx^j,
\label{interv2}
\ee
where  $\tilde\gamma_{ij}=\gamma_{ij}/a^2$ is a matrix with the
unit determinant.

The interval (\ref {interv1}) is similar formally to the ADM one
\cite{adm}, but with the lapse function defined as $N=a(1-\ptl _ m P ^ m) $, where $P ^ m $ is a
three-dimensional vector, and $\ptl_m$ is a conventional partial
derivative. Finally, restrictions $\ptl _ n(\ptl _ m N^m )=0$ and
$\ptl _ n(\ptl _ m P^m )=0$ arise on the Lagrange multipliers $\bm
N$ and $\bm P$ in FVT. The Hamiltonian $\mathcal H$ and momentum
$\mathcal P_i$ constraints in  the particular gauge $P^i=0$,
$N^i=0$  obey the constraint evolution equation \cite{Vesti}:
\bea
\ptl_\eta{\mathcal H}=\ptl_i\left(\tilde \gamma^{ij}\mathcal P_j\right),\label{5} \\
\ptl_\eta {\mathcal P_i}=\frac{1}{3}\ptl_i {\mathcal H},\label{6}
\eea
which admits adding of some constant to ${\mathcal H}$. Thus, the
constraint $\mathcal H$ is not necessarily to be zero, but
$\mathcal H=const$ is also allowed.

\section{Perturbations under Minkowski background}

\label{sec3}

Here we will consider an empty space-time filled only by vacuum and taking into account its quantum properties. The well-known
solution for an empty universe was found by Milne
\cite{Milne1935}. Although Milne himself does not use GR, from the
point of GR his universe represents closed empty universe
expanding linearly in cosmic time. Consideration of the mean
vacuum energy density and pressure in the framework of FVT gives a
a flat universe, which has a Milne's-like expansion stage
\cite{conf,Cherkas07,nonlin20} changed by the accelerated expansion.

Below, the scalar perturbations of the metric will be considered,
which look in the conformally-unimodular frame as \cite{Cherkas18}
\be
\fl ds^2=a(\eta,\bm
x)^2\left(d\eta^2-\left(\left(1+\frac{1}{3}\sum_{m=1}^3
\ptl_m^2F(\eta,\bm x)\right)\delta_{ij}-\ptl_i\ptl_jF(\eta,\bm
x)\right)dx^idx^j\right),
\label{int1}
\ee
where the perturbations of the locally defined scale factor 
\be
a(\eta,\bm x)=e^{\alpha(\eta)}(1+\Phi(\eta,\bm x)),
\label{a}
\ee
are expressed through a gravitational potential $\Phi$. A
stress-energy tensor could be written in the hydrodynamic
approximation \cite{Dodelson2003}
\be
T_{\mu\nu}=(p+\rho)u_{\mu}u_{\nu}-p\, g_{\mu\nu}.
\label{tmn}
\ee
The perturbations of the energy density $\rho(\eta,\bm
x)=\rho_v+\delta \rho(\eta,\bm x)$ and pressure $p(\eta,\bm
x)=p_v+\delta p(\eta,\bm x)$ will be considered around the vacuum
mean values, where the index $v$ will denote an uniform component
of the vacuum energy density and pressure.

The zero-order equations for a flat universe take the form
\cite{conf,Cherkas07,nonlin20}
\bea
M_p^{-2} e^{4 \alpha } \rho_v -\frac{1}{2} e^{2 \alpha}
\alpha ^{\prime 2}=const,\label{f1}\\
 \alpha ''+ \alpha^{\prime
2}=M_p^{-2} e^{2 \alpha} (\rho_v -3 p_v),\label{f2}
\eea
where $\alpha(\eta)=\log a(\eta)$.
 Here and
everywhere further, the system of units $\hbar=c=1$ is used as well
as the reduced Planck mass $M_p=\sqrt{\frac{3}{4\pi G}}$ is
implied. According to FVT \cite{Vesti}, the first Friedmann
equation (\ref{f1}) is satisfied up to some constant, and the main
parts of the vacuum energy density and pressure
\bea
\rho_v\approx (N_{boson}-N_{ferm})\frac{k_{max}^4}{16\pi^2
a^4},\label{rhvac}
\\
p_v=\frac{1}{3}\rho_v
\label{pvac}
\eea
do not contribute to the universe expansion. In the formula
(\ref{rhvac}), the  UV  cut-off and the number of bosonic and fermionic   degrees of freedom of the quantum fields appear because the zero-point
stress-energy tensor is an additive quantity \cite{Visser18}.
Here, we do not consider the supersymmetry hypotheses
\cite{Visser2019} due to the absence of evidence of the
supersymmetric particles to date.

Other contributors to the vacuum energy density are the terms
depending on the derivatives of the universe expansion rate
\cite{Cherkas07,nonlin20}. They  have the right order of
$\rho_{v}\sim M_p^2 H^2$, where $H$ is the Hubble constant, and
allow explaining the accelerated expansion of the universe. Then, the
energy density and pressure are
\cite{Cherkas07,nonlin20}:
\begin{equation}
\rho_v=\frac{a^{\prime
2}}{2a^6}M_p^2(2+N_{sc}){\mathcal S}_0,~~~~p_v=\frac{M_p^2(2+N_{sc}){\mathcal S}_0}{a^6}\left(\frac{1}{2}a^{\prime
2}-\frac{1}{3}a^{\prime\prime}a\right),
\label{eqn:rhoandp}
\end{equation}
where, $
    {\mathcal S}_0 = \frac{k_{max}^2}{8 \pi^2 M_p^2}.$
    Eqs. (\ref{eqn:rhoandp}) include the number of minimally coupled scalar fields
$N_{sc}$ plus two, because the gravitational waves give two
additional degrees of freedom \cite{Cherkas07}, whereas 
massless fermions and photons do not contribute to (\ref{eqn:rhoandp}) \cite{Cherkas07}.

The residual vacuum energy density and pressure (\ref{eqn:rhoandp})  lead to the accelerated universe expansion, which allows finding 
a momentum UV cut off
\be
k_{max}\approx \frac{12 M_p}{\sqrt{2+N_{sc}}}.
\label{kmax}
\ee
from the experimental value of the universe decceleration parameter \cite{Cherkas07,nonlin20,Haridasu2020}.

In this paper, we are interested in the local properties of a vacuum. Without including a real matter, if the constant in Eq. (\ref{f1})
compensates vacuum energy (\ref{rhvac}) exactly, one comes to
the static Minkowski space-time. Further, we will consider the
perturbations \cite{Cherkas18} under this background and set $\alpha(\eta)=0$ in
(\ref{a}).

\begin{figure}[ht]
\vspace{0. cm} \hspace{1 cm}
\includegraphics[width=15cm]{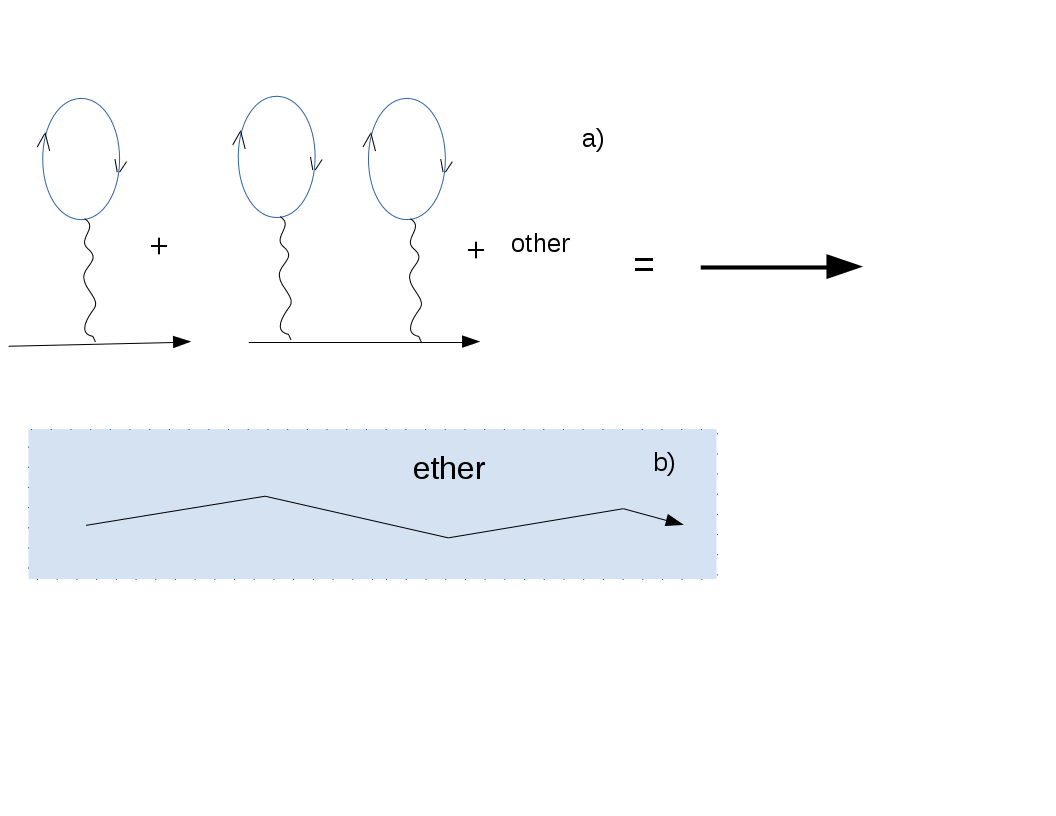}
\vspace{-3.0 cm} \caption{Illustration of vacuum influence to the
particle propagation a) in the QFT, where the vacuum loops
renormalize mass and charge of a particle, but do not prevent its
free motion b) and in the QG, where the space is filled by ether
due to the absence of a vacuum state.}
\label{vv}
\end{figure}

Generally, a vacuum can be considered as some fluid, i.e.,
``ether'' \cite{nonlin20}, but with some stochastic properties
among the elastic ones. Let us return to the stress-energy tensor
(\ref{tmn}) and introduce other variables
\bea
\wp(\eta,\bm x)=a^4(\eta,\bm x)\rho(\eta,\bm x), \label{var1}\\
\Pi(\eta,\bm x)=a^4(\eta,\bm x)p(\eta,\bm x)\label{var2}
\eea
for the reasons which will be explained below. The perturbations
around the uniform values can be written now as $\wp(\eta,\bm
x)=\rho_v+\delta\wp(\eta,\bm x)$, $\Pi(\eta,\bm
x)=p_v+\delta\Pi(\eta,\bm x)$. The vacuum-ether 4-velocity $u$ is
represented in the form of
\be
\fl u^{\mu}=\{(1-\Phi(\eta,\bm x)),\bm \nabla \frac{v(\eta,\bm
x)}{\wp(\eta,\bm x)+\Pi(\eta,\bm x)}\}\approx\{(1-\Phi(\eta,\bm
x)),\bm \nabla \frac{v(\eta,\bm x)}{\rho_v+p_v}\},
\label{vr}
\ee
where $v(\eta,\bm x)$ is a scalar function. Expanding all
perturbations into the Fourier series $\delta \wp(\eta,\bm
x)=\sum_{\bm k}\delta\wp_{\bm k}(\eta)e^{i \bm k \bm x }...$ etc.
results in the equations for the perturbations:

\bea
-6 \hat \Phi_{\bm k}'+k^2 \hat F_{\bm k}'+\frac{18}{M_p^2}
 \hat v_{\bm k}
=0,\label{con1}\\
 -6 k^2\hat \Phi_{\bm k}
+k^4\hat F_{\bm k} +\frac{18}{M_p^2}   \delta \hat\wp_{\bm k}\,
=0,\label{con2}\\
-12 \hat \Phi_{\bm k}-3 \hat F_{\bm k}''+k^2
\hat F_{\bm k}=0,\label{18}\\
-9 \hat \Phi_{\bm k}''-9 k^2 \hat \Phi_{\bm k} +k^4
\hat F_{\bm k} -\frac{9}{M_p^2}\left( 3 \delta \hat \Pi_{\bm k}-\delta \hat \wp_{\bm k}\right)=0,\label{eq4}\\
-\delta \hat \wp_{\bm k}'+k^2 \hat v_{\bm k}=0,\label{eq5}\\
\delta \hat \Pi_{\bm k}+\hat v_{\bm k}'=0.\label{lasteq}
\eea
It is remarkable that the choice of the variables (\ref{var1}),
(\ref{var2}), (\ref{vr}) means that the values $\rho_v$ and $p_v$
do not appear in the system (\ref{con1})-(\ref{lasteq}). The
second point is that the continuity and the Newton second law
equations (\ref{eq5}), (\ref{lasteq}) do not contain metric
perturbation.
 From now we will begin to consider the perturbation in
Eqs. (\ref{con1})-(\ref{lasteq}) as operators by writing a ``hat"
under every quantity. Here, we do not suppose the strong
nonlinearity \cite{Carlip2019} and assume a smallness of the
quantum fluctuations of space-time in this particular conformally
unimodular metric. Let us emphasize that the system
(\ref{con1})-(\ref{lasteq}) for a perturbation evolution is exact
in the first order on perturbations.  However, it is not closed.
To obtain a closed system, one needs, for instance, to specify the
sound speed for a perturbation of pressure. Still,
alternatively, as an approximation, we could strictly calculate pressure
and energy density by using the field theory under
unperturbed Minkowski space-time. Expressing $F_{\bm k}$ from Eq.
(\ref{con2}) and substituting it into Eq. (\ref{eq4}) leads to
\be
\hat \Phi_{\bm k}^{\prime\prime}+\frac{1}{3}k^2\hat \Phi_{\bm
k}+\frac{1}{M_p^2}\left(3\delta\hat\Pi_{\bm k}+\delta \hat\wp_{\bm
k}\right)=0.\label{single}
\ee
Although, generally, a gravity
causing an arbitrary curved space-time background does not allow a
well-defined and covariant vacuum state \cite{Birrell82}, we will approximately consider an operator $3\delta\hat\Pi_{\bm
k}+\delta \hat\wp_{\bm k}$  by using the creation and annihilation
operators under the Minkowski space-time background. Such an
approximation allows closing the system
(\ref{con1})-(\ref{lasteq}).  Nevertheless, let us
point out the difference between the quantum field theory (QFT)
and QG. As is shown in Fig.\ref{vv}, a test particle moves
straightforwardly in QFT. In a framework of the QG \cite{Kief}, the particle
has to undergo interaction with ether.

\subsection{Quantum fields as a source for energy density and  pressure perturbations}

Let us consider a single scalar field as an example of a quantum
field. Energy density and pressure of the scalar field in the pure
Minkowski space-time (without metric perturbation) have the form
\cite{Visser18}
\bea
 \hat p(\eta,\bm x)=\frac{\hat\varphi^{\prime 2}}{2}-\frac{(\bm \nabla
\hat\varphi)^2}{6},\\
\hat \rho(\eta,\bm x)=\frac{\hat\varphi^{\prime 2}}{2}+\frac{(\bm
\nabla \hat\varphi)^2}{2}.
\eea
All the  quantities may be expanded into the Fourier series $\hat
\varphi(\eta,\bm x)=\sum_{\bm k}\hat \phi_{\bm k}(\eta)e^{i\bm
k\bm x}$, $\hat p(\eta,\bm x)=\sum_{\bm k}\hat p_{\bm
k}(\eta)e^{i\bm k\bm x}$, where $\hat p_{\bm k}(\eta)=\int \hat
p(\eta,\bm x)e^{-i\bm k \bm x}d\bm x$, etc. For $\bm k\ne 0$, the
approximate identifying $\delta \hat \Pi_{\bm k}=\hat p_{\bm k}$
and $\delta \hat \wp_{\bm k}=\hat \rho_{\bm k}$ results in
\bea
\delta \hat \Pi_{\bm k}=\sum_{\bm q} \frac{1}{2}\hat\phi_{\bm
q}^{+\prime}\hat\phi_{\bm q+\bm k}^\prime -\frac{1}{6}(\bm q+\bm
k)\bm q\,\hat\phi_{\bm q}^+\hat\phi_{\bm q+\bm k},
\label{opp}\\
\delta \hat \wp_{\bm k}=\sum_{\bm q} \frac{1}{2}\hat\phi_{\bm
q}^{+\prime}\hat\phi_{\bm q+\bm k}^\prime+\frac{1}{2}(\bm q+\bm
k)\bm q\,\hat\phi_{\bm q}^+\hat\phi_{\bm q+\bm k},
\eea
so that the quantity $3\delta \hat \Pi_{\bm k}+\delta \hat
\wp_{\bm k}$ from Eq. (\ref{single}) is reduced to
\be
3\delta \hat \Pi_{\bm k}+\delta \hat \wp_{\bm k}=2\sum_{\bm q}
\hat\phi_{\bm q}^{+\prime}\hat\phi_{\bm q+\bm
k}^\prime.\label{quant}
\ee
Writing quantized field explicitly with creation and annihilation
operators \cite{Birrell82}
\be
\hat \phi_{\bm k}(\eta)=\frac{1}{\sqrt{2\omega_k}}\left(\hat
{\mbox{a}}^+_{-\bm k}e^{i\omega_k \eta}+\hat {\mbox{a}}_{\bm k}
e^{-i\omega_k\eta}\right),
\label{phiop}
\ee
allows obtaining from the Eqs. (\ref{quant}) and (\ref{phiop})
\bea
\fl 3\,\delta \hat \Pi_{\bm k}+\delta \hat \wp_{\bm k}=\sum_{\bm
q}\sqrt{\omega_q\omega_{|\bm q+\bm k|}}\biggl(
\hat{\mbox{a}}_{-\bm q}\hat{\mbox{a}}^+_{-\bm q-\bm
k}e^{i(\omega_{|\bm q+\bm k|}-\omega_q)\eta}+\hat{\mbox{a}}^+_{\bm
q}\hat{\mbox{a}}_{\bm q+\bm k}e^{i(\omega_q-\omega_{|\bm q+\bm
k|})\eta}\nonumber\\~~~~~~~~~~~~~~~~~~~~~~~~- \hat{\mbox{a}}_{-\bm
q}\hat{\mbox{a}}_{\bm q+\bm k}e^{-i(\omega_{|\bm q+\bm
k|}+\omega_q)\eta}-\hat{\mbox{a}}^+_{\bm q}\hat{\mbox{a}}^+_{-\bm
q-\bm k}e^{i(\omega_{|\bm q+\bm k|}+\omega_q)\eta}\biggr),~~
\label{delp}
\eea
where for a massless scalar field $\omega_{\bm k}=|\bm k|$. As is
seen from Eq. (\ref{delp}), the perturbations have the general
form:
\be
 3\,\delta \hat \Pi_{\bm k}+\delta \hat \wp_{\bm
k}=\sum_m \hat {\mathcal P}_{m\bm k}e^{i\Omega_{m\bm k}\eta},
\ee
where the frequencies $\Omega_{m\bm k}$ take the values of
$\omega_q-\omega_{|\bm q+\bm k|}$, $-\omega_q+\omega_{|\bm q+\bm
k|}$, $\omega_q+\omega_{|\bm q+\bm k|}$ and
$-\omega_q-\omega_{|\bm q+\bm k|}$.  That allows finding the
solution of Eq. (\ref{single}) as

\be
\hat \Phi_{\bm k}(\eta)=-\frac{1}{M_p^2}\sum_m\frac{\hat {\mathcal
P}_{m\bm k} e^{i\Omega_{m\bm k}\eta}}{\Omega^2_{m\bm k}-k^2/3}.
\label{genf}
\ee

Using Eqs. (\ref{delp}) and (\ref{genf}), the final expression for
the metric perturbation $\hat \Phi_{\bm k}(\eta)$ acquires the
form
\bea
\fl\hat \Phi_{\bm k}(\eta)=\frac{1}{M_p^2}\sum_{\bm
q}\sqrt{\omega_q\omega_{|\bm q+\bm k|}}\Biggl(\frac{1}
{(\omega_{|\bm q+\bm
k|}+\omega_q)^2-k^2/3}\biggl(\hat{\mbox{a}}_{-\bm
q}\hat{\mbox{a}}_{\bm
q+\bm k}e^{-i(\omega_{|\bm q+\bm k|}+\omega_q)\eta}+\nonumber\\
\fl\hat{\mbox{a}}^+_{\bm q}\hat{\mbox{a}}^+_{-\bm q-\bm
k}e^{i(\omega_{|\bm q+\bm k|}+\omega_q)\eta}\biggr)- \frac{1}
{(\omega_{|\bm q+\bm k|}-\omega_q)^2-k^2/3}
\biggl(\hat{\mbox{a}}_{-\bm q}\hat{\mbox{a}}^+_{-\bm q-\bm
k}e^{i(\omega_{|\bm q+\bm
k|}-\omega_q)\eta}+\nonumber\\\hat{\mbox{a}}^+_{\bm
q}\hat{\mbox{a}}_{\bm q+\bm k}e^{i(\omega_q-\omega_{|\bm q+\bm
k|})\eta}\biggr) \Biggr).~
\label{finA}
\eea
The most interesting parameter is a correlator:
\be
\fl <0|\hat \Phi(\eta,\bm x)\hat \Phi(\tau,\bm
x^\prime)|0>=\chi(\tau-\eta,|\bm x-\bm
x^\prime|),~~~~~\chi(\tau-\eta,x)=\sum_{\bm k}
S(\tau-\eta,k)e^{i\bm k\bm x},
\label{potr}
\ee
which determines fluctuations of the gravitational potential $\hat
\Phi(\eta,\bm r)$ in a vacuum state defined for the creation and
annihilation operators. An explicit formula for $S(\eta-\tau,k)$
looks as
\bea
\fl S(\tau-\eta,k)=<0|\hat \Phi_{\bm k}^+(\eta)\hat \Phi_{\bm
k}(\tau)|0>=\frac{18}{(2\pi)^3M_p^4}\int\frac{e^{i(\tau-\eta)(\omega_{\bm
q}+\omega_{\bm q+\bm k})} \omega_{\bm q} \omega_{\bm k+\bm q}
d^3\bm q}{\left(k^2-3( \omega_{\bm q}+
   \omega_{\bm k+\bm q})^2\right)^2 },
   \label{sigm2}
   \eea
where the summation over $\bm q$ has been changed by the
integration as $\sum_{\bm q}\rightarrow\frac{1}{(2\pi)^3} \int
d^3\bm q$. To calculate this integral, the spherical coordinates
can be applied,
 in which
$\omega_{\bm k+\bm q}=\sqrt{k^2+2k q \cos \theta+q^2}$,
 $~~\bm q(\bm q+\bm k)=q^2+k q \cos\theta$, $~~d^3\bm q=2\pi
 q^2d q\sin\theta d\theta$.
It is more convenient to calculate a spectral function $\tilde S(\omega,\bm
q)$ of the correlator (\ref{sigm2})
\bea
\fl \tilde S(\omega,k)=\frac{1}{2\pi}\int S(\eta,k)e^{-i\omega
\eta}d\eta=\frac{18}{(2\pi)^3M_p^4}\int_{q<k_{max}}\frac{\delta(\omega_{\bm
q}+\omega_{\bm q+\bm k}-\omega) \omega_{\bm q} \omega_{\bm k+\bm
q} d^3\bm q}{\left(k^2-3( \omega_{\bm q}+
   \omega_{\bm k+\bm q})^2\right)^2 }=\nonumber\\
\fl\left\{\begin{array}{c}
 \frac{1}{160 \pi ^2 {M_p}^4}\left(5+\frac{4 k^4}{\left(k^2-3 \omega ^2\right)^2}\right), \,q<\omega<2k_{max} \\
 0,  \,\,\mbox{otherwise}\end{array}\right.\approx \left\{\begin{array}{c}
 \frac{1}{32 \pi ^2 {M_p}^4}, \,q<\omega<2k_{max} \\
 0,  \,\,\mbox{otherwise}.
\end{array}\right.
\label{spectrum}
\eea

Taking into account that  the main contribution originates from
large $q$, one could also calculate simultaneous correlator 
\be
\fl <0|\hat \Phi(\eta,\bm x)\hat \Phi(\eta,\bm x^\prime)|0>\approx
\frac{  k_{max}}{4(2\pi)^2M_p^4 }\delta(\bm x-\bm x^\prime),
\label{cont}
 \ee
which corresponds to the contact interaction and was
used in  \cite{prev}. However, more careful
analysis based on \ref{app3} shows that using of (\ref{cont}) is insufficient 
 and the spectral function (\ref{spectrum}) of the non-simultaneous correlator
 (\ref{potr})
 plays a
role.

 \section{Massive particle in a random medium}

 \subsection{Point particles}

Let us first consider nonrelativistic point massive particles propagating
among the fluctuations of the gravitational potential \cite{Vacchini2009}. The evolution
of a system could be described by the Fokker-Plank type equation
given in \ref{app3}
\be
\fl \ptl_\eta f_{\bm k}(\bm p)+i(E_{{\bm p}+{\bm k}/2}-E_{{\bm
p}-{\bm k}/2})f_{\bm k}(\bm p)= -i\,K_1\,\bm k\frac{\ptl f_{\bm
k}}{\ptl \bm p} +2i K_{2}\,\bm k\bm p\,\Delta_{\bm p}f_{\bm k}(\bm
p) +2 i K_{3}\,p_i k_j \frac{\ptl^2 f_{\bm k}}{\ptl p_j\ptl
p_i},~~
\label{fokker}
\ee
where  $\Delta_{\bm p}$ is a Laplacian over $\bm p$, the constants $K_1=\frac{m^2 N_{all}}{32 \pi^2M_p^4}\tilde K_1,
~K_2=\frac{m^2 N_{all}}{32 \pi^2M_p^4}\tilde K_2, ~...$, and
$\tilde K_1, \tilde K_2, \tilde K_3$ are given in \ref{app3}.  In
the difference from (\ref{rhvac}), the quantities $K_i$ contain
sum $N_{all}=N_{boson}+N_{ferm}$ of the bosonic and fermionic
degrees of freedom, because correlator (\ref{sigm2}) is the second
order on gravitational potential $\Phi$, whereas $\Phi$ is
proportional to the energy density and pressure according to
 Eq.
(\ref{single}).

It is suggested that the Fokker-Planck equation is applicable for
particles of large mass when the momentum of a particle is larger
than the maximal momentum transferred, which is considered to be of the
order of $M_p$ for point-like particle. Since the Migdal equation (\ref{mm}) is too complicated for solution, the Fokker-Planck equation could be used to obtain an estimation of decoherence for particles of a smaller mass.

Due to the smallness of the right-hand side of (\ref{fokker}), one
needs to find a solution only in the first order on the constants
$K_1$, $K_2$, $K_3$. For this aim, it is sufficient to substitute
approximate solution (\ref{ff}) into the right-hand side of
(\ref{fokker}) and then solve it. This gives
\bea
\fl f_{\bm k}(\bm p,\eta)\approx \tilde f_{\bm k}(\bm
p,\eta)\Biggl(1+K_1\biggl(-\frac{{k^2}\eta^2}{2 m}+\frac{2 i \eta \bm k
({\bm p}-{\bm p_0})}{\Gamma ^2}\biggr)+\nonumber\\\fl
K_2\biggl(\frac{2 \eta {\bm k\bm p}  \left(-i \Gamma ^4 {k^2} \eta^2-6
\Gamma ^2 m (\bm k ({\bm p}-{\bm p_0})\eta+3
   i m)+12 i m^2 {(\bm p-\bm p_0)^2}\right)}{3 \Gamma ^4 m^2}\biggr)+\nonumber\\
   \fl~~~~~ K_3\biggl(-\frac{2 \eta \left(\eta \left(-{(\bm k\bm p_0)}
   {(\bm k\bm p)}+{(\bm k\bm p)}^2+{k^2}
   ({p^2}- {\bm p\bm p_0})\right)+2 i m {\bm k\bm p}\right)}{\Gamma ^2 m}+\nonumber\\\frac{8 i \eta
   ({\bm k\bm p}-{\bm k\bm p_0} ) ({p^2}-{\bm p_0} {\bm p})}{\Gamma ^4}-\frac{2 i
   {(\bm k\bm p)} {k^2} \eta^3}{3 m^2}\biggr)\Biggr),
   \label{resf}
\eea
where $\tilde f_{\bm k}(\bm p,\eta)$ is given  in \ref{app1} by (\ref{ff}).
Substituting the solution (\ref{resf}) into (\ref{dec}) gives in the
first order on the constants $K_1$, $K_2$, $K_3$
\be
\int  f_{\bm k}(\bm p,\eta)f_{-\bm k}(\bm p,\eta)d^3\bm pd^3\bm k\approx
1-(3 {K_1}+3 {K_2}+6 {K_3})\frac{\Gamma ^2 \eta^2 }{m}.
\label{resdec}
\ee
As one could see, the interaction with vacuum produces decoherence
expressed in the decreasing of a "purity" (\ref{dec}) of a
particle state according to (\ref{resdec}). From Eq.
(\ref{resdec}), the decoherence time is estimated as
\be
t_{dec}\approx\frac{1}{\Gamma}\sqrt{\frac{m}{3 {K_1}+3 {K_2}+6
{K_3}}}.
\label{tdec}
\ee

\begin{figure}[th]
  \includegraphics[width=11cm]{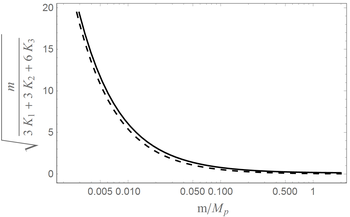}\\
  \caption{
   A  dimensionless quantity determining decoherence time and length by Eq. (\ref{tdec}) (solid line),  an approximation $\sqrt{\frac{m}{3 {K_1}+3 {K_2}+6 {K_3}}}\sim \frac{4 M_p}{3\sqrt{3\,N_{all}}\pi\,m}$ (dashed line).
  
  It is taken $N_{sc}=4$ and $N_{all}=126$ in Eq. (\ref{kmax}) for $k_{max}$.} \label{fig2}
\end{figure}

It is convenient to measure the decoherence length
$L_{dec}=t_{dec}\mathcal V$ in terms of the localization length
$1/\Gamma$ of the wave packet (see \ref{app1}). Particle velocity is defined as $\mathcal V=p_0/m$. Dependence of the constant $\sqrt{\frac{m}{3
{K_1}+3 {K_2}+6 {K_3}}}$ is shown in Fig. \ref{fig2}, where also
an approximate
 expression  is shown. Using this
 approximate expression, one comes to
$ L_{dec}\approx \frac{4
M_p}{3\sqrt{3 \,N_{all}}\pi\,m}\frac{\mathcal V}{\Gamma}. $
That is, a point-like particle of mass
$ m\sim \frac{4 M_p\mathcal V}{3\sqrt{3  N_{all}}\pi}$ loses coherence at a distance equal to the length of the
wave packet $1/\Gamma$. It should be noted that interaction with
the ether does not produce particle scattering because the momentum
distribution $f_0(\bm p)$ does not change, nevertheless the
decoherence arises.

\subsection{Particles of a finite size}

A real particle of a large mass has a finite size, which
restricts momentums transferred by the form factor. Approximately,
momentum transferred $q$ in the Eqs. (\ref{b1}), (\ref{b2}) should
be restricted by $q<1/d$, where $d$ is size of a particle. In this
case, the calculation of the integrals gives $
K_1=\frac{ m\,N_{all}}{192\pi^2(M_p d)^4}$, $K_2=\frac{N_{all}}{1200\pi^2
M_p^4d^5}$, $K_3=\frac{N_{all}}{2400\pi^2 M_p^4d^5}$. The main
contribution to the decoherence length  for large mass particles
originates from the constant $K_1$ and gives
\be L_{dec}\approx
\frac{\mathcal
V}{\Gamma}\sqrt{\frac{m}{3K_1}}\approx\frac{8\pi (M_p
d)^2}{\sqrt{N_{all}}} \frac{\mathcal V}{\Gamma}.\ee
 This quantity seems very large
and unobservable in a matter-wave interferometry
\cite{Hornberger2004,Nature,Braz,walter2016coherence,Brand2020,brand2017matterwave}, because increasing of the particle mass does not
decrease decoherence length.  On the other hand, the large mass
particles usually have internal degrees of freedom and another
decoherence mechanisms
\cite{Pikovski2015,Gooding2014,Gooding2015,Ralph2014,Plato2016,Pang16,Carlesso2016,Dehda17}
related with these internal degrees of freedom works.
Also, a particle spin could be considered as an internal degree of freedom and, thereby, produce decoherence 
\cite{Esfahani07,Ahmadi14}. 

It should also be noted that another branch of combining gravity
and quantum mechanics exists, namely, reduction of the wave
function due to gravitational interaction \cite{penr}. That is
beyond an ``usual" QG and the content of this paper.

\section{Discussion and conclusion}

The QG must produce a considerable decoherence effect for a pure problem formulated for the point-like massive particles. Interestingly, this effect originates not due to the on-shell multiple scattering
\cite{bar,Mig1}, forbidden by the energy conservation, but from the off-shell effects.

For the real particles of finite size, the form factor restricts
the momentum transferred. That reduces the effect of QG decoherence to an unobservable level. This QG decoherence could not compete with
the decoherence arising from the interaction of the internal degrees
of freedom of a composite particle with the gravitational field.

A general remark about the decoherence should also be made.
From the point of quantum gravity, the universe as a whole exists
in a single quantum state \cite{cherkas2020cosmological} and has
zero entropy. Consequently, one could not consider a massive
particle as completely isolated because it is always
embedded into the general quantum state. Thus, any object does not
lose its quantum properties but becomes more entangled with the
universe's general quantum state.

It is of interest to analyze QG vacuum effects on the propagation
of high energy gamma quanta  \cite{Magic,Amenomori19} in the
universe. We plan to perform this investigation in the nearest
future because the preliminary analysis \cite{prev} based on a simultaneous correlator of the gravitational potential is insufficient. 

\section*{Acknowledgments}
VLK acknowledges the funding from the European Union Horizon 2020
research and innovation program under the Marie Sklodowska-Curie
Grant No. 713694 (MULTIPLY).

\appendix
\section{Quantum mechanical evolution of the wave packet }
\label{app1}

  A momentum wave
packet of a freely moving particle can be written as
\be
\psi(p,t)=\psi_0(p)e^{-i\frac{p^2}{2m}t},
\ee
because it obeys the Schr\"{o}dinger equation
\be
i\frac{\ptl \psi}{\ptl t}=\frac{p^2}{2m}\psi.
\ee
The function $f_{\bm k}(\bm p)$ corresponding to this  pure
quantum state is
\be
f_{\bm k}(\bm p)=\psi(\bm p+{\bm k}/{2})\psi^*(\bm p-{\bm k}/{2}).
\ee
For the Gaussian wave packet $\psi(\bm p,
t)=\pi^{-3/4}\Gamma^{-3/2}\,e^{-\frac{(\bm p-\bm
p_0)^2}{2\Gamma^2}-i\frac{p^2}{2m}t}$, the function $f_{\bm k}(\bm
p)$ takes the form of
\be
f_{\bm k}(\bm p)=\pi^{-3/2}\Gamma^{-3}\,e^{-\frac{(\bm p-\bm
p_0)^2}{\Gamma^2}-\frac{k^2}{4\Gamma^2}-i\frac{\bm k\bm p}{m}t}.
\label{ff}
\ee
For pure states, the density matrix $\rho_{{\bm p} {\bm
p}^\prime}=\frac{1}{(2\pi)^3}f_{{\bm p}-{\bm
p}^\prime}\left(\frac{{\bm p}^\prime+\bm p}{2}\right)$  satisfies
\cite{lan3} $ \sum_{ {\bm p}^\prime}\rho_{{\bm p} {\bm
p}^\prime}\rho_{{\bm p}^{\prime} {\bm
p}^{\prime\prime}}=\rho_{{\bm p} {\bm p}^\prime} $ or $ \sum_{
{\bm p},{\bm p}^\prime}\rho_{{\bm p} {\bm p}^\prime}\rho_{{\bm
p}^{\prime} {\bm p}}=1 $. The last equality expressed in terms of the
function $f_{\bm k}(\bm p)$ as
\be
\int  f_{\bm k}(\bm p)f_{-\bm k}(\bm p)d^3\bm pd^3\bm k=1
\label{dec}
\ee
could serve as a criterion of "purity" of a system state.

\section{Wigner function evolution in a random medium}
\label{app3}

Let us consider an equation for the density matrix $\wp$
\be
i\ptl_\eta \hat \wp=[\hat H_1+H_2+\hat V,\hat\wp], \label{inrho}
\ee
where operators $\hat H_1$, $\hat H_2$ describe the test particle and
ether-medium, respectively \cite{Mig,Mig1,Kag}.  The operator $\hat V(\bm x)$ is an operator acting in a test particle Hilbert space described by $\bm x$, and, besides, acting to the vacuum-ether variables. We will omit hats everywhere further in this Appendix.
Let us introduce the density matrix of a particle by the averaging
$\rho=Tr_2\wp$, then it satisfies the equation
\be
i\ptl_\eta \rho=[H_1,\rho]+Tr_2[V,\wp].
\label{rho1}
\ee
The formal solution of Eq. (\ref{inrho}) could be written as
\be
\wp(\eta)=-i\int_{-\infty}^\eta
e^{i(H_1+H_2)(\tau-\eta)}[V,\wp(\tau)]e^{-i(H_1+H_2)(\tau-\eta)}d\tau.
\ee
This expression can be  substituted into the Eq. (\ref{rho1})
and one comes to
\be
\fl i\ptl_\eta\rho=[H_1,\rho]-i Tr_2\int_{-\infty}^\eta[V
,e^{i(H_1+H_2)(\tau-\eta)}[V,\wp(\tau)]e^{-i(H_1+H_2)(\tau-\eta)}]d\tau.
\ee
For further approximation, the density matrix is factorized as $\wp(\tau)=\rho(\tau)\rho_2(\tau)$. Then, one has to take into account that the calculations of the correlator of the interaction in the Sec. \ref{sec3} have been performed in the Heisenberg picture over medium-ether variables. Thus, we have to put interaction into the Heisenberg form using $V=e^{-iH_2\tau}V(\tau) e^{iH_2\tau}$, and, respectively bring the density matrix of a vacuum-medium into to the static form by writing $\rho_2(\tau)=e^{-iH_2\tau}\rho_2
 e^{iH_2\tau}$. This leads to the equation
 \be
\fl i\ptl_\eta\rho=[H_1,\rho]-i Tr_2\int_{-\infty}^\eta[V(\eta)
,e^{iH_1(\tau-\eta)}[V(\tau),\rho(\tau)\rho_2]e^{-i
H_1(\tau-\eta)}]d\tau.
 \ee
Else, in terms of the matrix elements corresponding to the plane
waves \cite{Mig,Kag}:
\bea
\fl i\ptl_\eta\rho_{{\bm p} {\bm p}^\prime}=(E_{\bm p}-E_{{\bm
p}^\prime})\rho_{{\bm p}{\bm p}^\prime}-i\sum_{\bm q,{\bm
q}^\prime}\int_{-\infty}^\eta\biggl( <V_{{\bm p}-{\bm
q}}(\eta)V_{{\bm q}-
{\bm q}^\prime}(\tau)>\rho_{{\bm q}^\prime {\bm p}^\prime}(\tau)
e^{i(E_{\bm q}-E_{{\bm p}^\prime})(\tau-\eta)}-\nonumber\\
\fl <V_{{\bm p}-\bm q}(\eta)V_{{\bm q}^\prime -{\bm
p}^\prime}(\tau)>\rho_{\bm q {\bm q}^\prime}(\tau)e^{i(E_{\bm
q}-E_{{\bm p}^\prime})(\tau-\eta)}-<V_{{\bm p}-{\bm
q}}(\tau)V_{{\bm q}^\prime -{\bm p}^\prime}(\eta)>\rho_{{\bm q}
{\bm q}^\prime}(\tau)e^{i(E_{\bm p}-E_{{\bm q}^\prime})(\tau-\eta)}+\nonumber\\
\rho_{{\bm p}\bm q}(\tau)<V_{\bm q -{\bm q}^\prime}(\tau)V_{{\bm
q}^\prime - {\bm p}^\prime}(\eta)>e^{i(E_{\bm p}-E_{{\bm
q}^\prime})(\tau-\eta)} \biggr)d\tau,
\label{elems}
\eea
where an averaging $<>$ implies
\[
 <V_{{\bm q}}(\eta)V_{ {\bm p}}(\tau)>=Tr_2\biggl(V_{\bm q}(\eta)V_{ {\bm
p}}(\tau)\rho_2\biggr)\equiv <0|V_{\bm q}(\eta)V_{ {\bm
p}}(\tau)|0>.
\]
According to (\ref{sigm2}), (\ref{spectrum}), one has:
\be
\fl <0|V_{{\bm q}^\prime}(\eta)V_{\bm q}(\tau)|0>=m^2\delta_{-{\bm
q}^\prime,\bm q}S(\bm q,\tau-\eta)=m^2\delta_{-{\bm q}^\prime,\bm
q}\int \tilde S(\bm q,\omega)e^{i\omega(\tau-\eta)}d\omega.
\ee
After changing the integration variable $\tau^\prime=\tau-\eta$ in
the integral (\ref{elems}) and using approximately
\be
\rho_{\bm p\bm q}(\eta+\tau^\prime)\approx e^{-i(E_{\bm p}+E_{\bm
q})\tau^\prime}\rho_{\bm p\bm q}(\eta)
\ee
in the second-order on the interaction  expression in the right-hand side of (\ref{elems}),
we come to the following equation:
\bea
\fl i\ptl_\eta\rho_{{\bm p} {\bm p}^\prime}=(E_{\bm p}-E_{{\bm
p}^\prime})\rho_{{\bm p}{\bm p}^\prime}-i m^2\sum_{\bm
q}\int\biggl( \tilde S(q,\omega)\biggl((\Delta(E_{\bm p+\bm
q}-E_{\bm
p}+\omega)+\nonumber\\
\fl\Delta(E_{\bm p^\prime}-E_{\bm p^\prime+\bm q}-\omega))
\rho_{{\bm p} {\bm p}^\prime}(\eta)-(\Delta(\omega-E_{{\bm
p}^\prime}+E_{{\bm p}^\prime+\bm q})+\nonumber\\
~~~~~~~~~~~~~~~~~~~~~~~~\Delta(-\omega+E_{\bm p}-E_{\bm p+\bm q}))\rho_{{\bm p}+{\bm
q},{\bm p}^\prime+\bm q}\biggr)d\omega,
\label{follow}
\eea
where
\be
\Delta(\omega)=\int_{-\infty}^0e^{i\omega\tau}d\tau=\pi\delta(\omega)-i\mathcal
P \frac{1}{\omega}
\label{delta}
\ee
contains the Dirac $\delta$-function and the main value generalized
function $ \mathcal P \frac{1}{\omega}$  \cite{vlad}.

 In terms of the Fourier transform  of the  Winger function \cite{Mig} $\rho_{{\bm p} {\bm
p}^\prime}=f_{{\bm p}-{\bm p}^\prime}\left(\frac{{\bm
p}^\prime+\bm p}{2}\right)$  or $f_{\bm k}(\bm p)=\rho_{\bm p+\bm
k/2,\bm p-\bm k/2}$ Eq. (\ref{follow}) is written as
\bea
\fl \ptl_\eta f_{\bm k}(\bm p)+i(E_{{\bm p}+{\bm k}/2}-E_{{\bm
p}-{\bm k}/2})f_{\bm k}(\bm p)= m^2\sum_{\bm q}\int\tilde
S(q,\omega)\biggl( \bigl(\Delta(\omega-E_{{\bm p}-{\bm
k}/2}+E_{{\bm p-{\bm k}/2+{\bm q}}})+\nonumber \\
\fl\Delta(-\omega+E_{{\bm p}+{\bm k}/2}-E_{{\bm p+{\bm k}/2+{\bm
q}}})\bigr)f_{\bm k}(\bm p+\bm q)- \bigl(\Delta(\omega+E_{{\bm
p}+{\bm k}/2+\bm q}-E_{{\bm p+{\bm k}/2}})+\nonumber \\
\fl\Delta(-\omega+E_{{\bm p}-{\bm k}/2}-E_{{\bm p-{\bm k}/2+{\bm
q}}})\bigr)f_{\bm k}(\bm p) \biggr)d^3 \bm q \,d\omega.
\label{mm}
\eea
Summation over $\bm q$ could be changed by the integration. Hence,
the Dirac delta-functions in (\ref{delta}) produce zero
contribution to the integral. Actually, since the minimal value of
$\omega$ is restricted by $q$ according to (\ref{spectrum}), the
value $\omega+\frac{(\bm p+\bm q)^2}{2m}-\frac{
p^2}{2m}>q+\frac{\bm p\bm q}{m}+q^2>0$ by the virtue of $p=m\mathcal V<m$.
Thus, only the second main value term in (\ref{delta}) gives
contribution.

 In a diffusion approximation, one may
expand $f_{\bm k}(\bm p+\bm q)-f_{\bm k}(\bm p)\approx\bm q\frac{\ptl
f_{\bm k}}{\ptl \bm p}+\frac{1}{2}q_i q_j\frac{\ptl^2 f_{\bm
k}(p)}{\ptl p_i\ptl p_j}$, and the following integrals arise
\bea
\fl \int_{q<k_{max}}\int_q^{2k_{max}}\left( \frac{\bm
q}{\omega+E(\bm p+\bm q)-E(\bm p)}-\frac{\bm q}{\omega+E(\bm
p^\prime+\bm q)-E(\bm
p^\prime)}\right)d\omega\, d^3\bm q  \approx\nonumber\\
-(\bm p-\bm p^\prime)\tilde K_1(m,k_{max})+O(p^4),
\label{b1}
\\
\fl \tilde K_1=\frac{4}{3} m \Biggl({k_{max}} (3 {k_{max}}+4 m)-8
k_{max}^{3/2}\sqrt{m} \,\arctan\left(\frac{1}{2}
   \sqrt{\frac{k_{max}}{m}}\right)+\nonumber\\~~~~~~~~~~~~~~~~~~~~~~~~8m^2 \ln
\left(\frac{2m}{k_{max}+2
   m}\right)\Biggr),
\nonumber
\label{int1a}
\\
\fl \int_{q<k_{max}}\int_q^{2k_{max}}\left( \frac{\bm q\otimes \bm
q}{E(\bm p+\bm q)-E(\bm p)+\omega}-\frac{\bm q\otimes \bm q}{E(\bm
p^\prime+\bm q)-E(\bm
p^\prime)+\omega}\right)d\omega \,d^3\bm q  \approx\nonumber\\
(p^2-p^{\prime 2})\tilde K_{2}(m,k_{max})\bm I+(\bm p\otimes\bm p-\bm
p^\prime\otimes\bm p^\prime)\tilde K_{3}(m,k_{max})+O(p^4),
\label{b2}
\eea
\bea
\fl \tilde K_{2}=\frac{8}{15} m \Biggl({k_{max}} \left({k_{max}} \left(\frac{2 {k_{max}} m}{{k_{max}}^2+6
  {k_{max}} m+8 m^2}+3\right)-10 \sqrt{{k_{max}} m} \arctan\left(\frac{1}{2}
   \sqrt{\frac{k_{max}}{{m}}}\right)+8 m\right)+\nonumber\\8 m^2 \left(5 \ln \left(\frac{2
   m}{{k_{max}}+2 m}\right)+6\, \mbox{arctanh} \left(\frac{{k_{max}}}{{k_{max}}+4
   m}\right)\right)\Biggr),
   \eea

\bea
\fl \tilde K_3=\frac{16}{15} m \Biggl({k_{max}} \Biggl({k_{max}} \left(\frac{2 {k_{max}} m}{{k_{max}}^2+6
   {k_{max}} m+8 m^2}+3\right)-10 \sqrt{m\,{k_{max}} } \arctan\left(\frac{1}{2}
   \sqrt{\frac{k_{max}}{{m}}}\right)+\nonumber\\\fl 8 m\Biggr)-
 4 m^2 \left(5 \ln \left(\frac{2
   m}{{k_{max}}+2 m}\right)+18\, \mbox{arctanh}\left(\frac{{k_{max}}}{{k_{max}}+4
   m}\right)\right)\Biggr).
\eea

Finally, we have:
\bea
\fl \ptl_\eta f_{\bm k}(\bm p)+i\frac{\bm p\bm k}{m}f_{\bm k}(\bm
p)= \frac{i\, m^2}{32\pi^2M_p^4}\Biggl(-\tilde K_1\,\bm
k\frac{\ptl f_{\bm k}(\bm p)}{\ptl \bm p}\nonumber\\ ~~~~~~~~~+\,2
\tilde K_{2}\,\bm k\bm p\,\Delta_{\bm p}f_{\bm k}(\bm p) +2\,\tilde
K_{3}\, p_i k_j \frac{\ptl^2 f_{\bm k}(\bm p)}{\ptl p_j\ptl
p_i}\Biggr).
\eea

\section*{References}

\bibliography{gravity}

\end{document}